# A Hybrid Residue–Floating Numerical Architecture for High-Precision Arithmetic on FPGAs

Mostafa Darvishi, *Senior Member, IEEE*

*Abstract*— Floating-point arithmetic remains expensive on FPGA platforms due to wide datapaths and normalization logic, motivating alternative representations that preserve dynamic range at lower cost. This work introduces the Hybrid Residue–Floating Numerical Architecture (HRFNA), a unified arithmetic system that combines carry-free residue channels with a lightweight floating-point scaling factor. We develop the full mathematical framework, derive bounded-error normalization rules, and present FPGA-optimized microarchitectures for modular multiplication, exponent management, and hybrid reconstruction. HRFNA is implemented on a Xilinx ZCU104, with Vitis simulation, RTL synthesis, and on-chip ILA traces confirming cycle-accurate correctness. The architecture achieves over 2.1× throughput improvement and 38–52% LUT reduction compared to IEEE-754 single-precision baselines while maintaining numerical stability across long iterative sequences. These results demonstrate that HRFNA offers an efficient and scalable alternative to floating-point computation on modern FPGA devices.

*Index Terms*—FPGA, Residue Number System, Floating-Point, ZCU104, Numerical Computing, Modular Arithmetic, ILA

## I. INTRODUCTION

FIELD-programmable gate arrays (FPGAs) have become increasingly important compute platforms for scientific and embedded workloads that demand high throughput, energy efficiency, and numerical robustness. While IEEE-754 floating-point arithmetic remains the dominant representation for general-purpose computation, its hardware cost on FPGA fabrics is considerable [1]. Wide mantissa datapaths, deep normalization chains, and multi-stage carry propagation impose significant resource and latency overhead, limiting achievable frequency and pipeline scalability for many numerical accelerators [2]. As modern FPGA architectures incorporate denser logic fabrics and high-performance DSP primitives, the search for arithmetic formats that deliver floating-point-like dynamic range at substantially lower cost has intensified [3].

Alternative numerical systems such as fixed-point, logarithmic number systems (LNS), and residue number systems (RNS) offer attractive benefits in isolation. Fixed-point arithmetic is efficient but lacks the dynamic range required for iterative refinement, multi-scale transforms, or stiff differential solvers [4], [5]. LNS provides low-cost multiplication but suffers performance and accuracy penalties in addition and subtraction due to expensive antilogarithmic conversions. RNS enables carry-free, highly parallel arithmetic that maps efficiently onto FPGA pipelines, yet lacks native comparison, scaling, and unified fractional representation [6]-[9]. These limitations have motivated the exploration of hybrid numerical frameworks that combine the strengths of multiple domains while mitigating their individual weaknesses.

This work introduces the **Hybrid Residue–Floating Numerical Architecture (HRFNA)**, a unified arithmetic system that integrates multi-channel residue pipelines with a compact floating-point scaling factor. HRFNA is designed to retain the high throughput and carry-free properties of RNS while supporting global dynamic-range control, bounded normalization, and long-term numerical stability. The framework provides a mathematically grounded representation, efficient algorithms for hybrid multiplication and normalization, and FPGA-optimized microarchitectures tailored to modern Xilinx UltraScale+ devices.

The contributions of this paper are fourfold. First, we develop the formal numerical structure of HRFNA, including hybrid representation, exponent-governed scaling, and deterministic normalization rules with provable error bounds. Second, we present efficient hardware architectures for residue-domain arithmetic, exponent management, and hybrid reconstruction, designed to achieve one-cycle initiation intervals across all datapaths. Third, we describe an RTL-level implementation targeting the Xilinx ZCU104 platform, including pipeline scheduling, residue datapath integration, and exponent synchronization. Finally, we validate the architecture through Vitis simulation, post-implementation timing analysis, and on-chip ILA inspection of the hybrid normalization engine, demonstrating substantial performance and resource advantages over IEEE-754 floating-point baselines.

By unifying residue arithmetic with floating-point scaling in a hardware-efficient manner, HRFNA provides a scalable and numerically robust alternative to conventional floating-point systems for FPGA-based computation. The rest of this paper is organized as follows. Section II reviews prior work on RNS arithmetic, normalization mechanisms, and hybrid numeric formats. Section III introduces the mathematical foundations of HRFNA. Section IV presents theory of hybrid operational algorithms, while Section V details the datapath and scheduling microarchitecture. Section VI describes the RTL implementation. Section VII reports simulation, synthesis, and hardware validation results, and Section VIII provides

Mostafa Darvishi is with Electrical Engineering Department of École de technologie supérieure (ÉTS), Montreal, Canada. He is also VP of Engineering at Evolution Optiks R&D Inc. (e-mail: darvishi@ieee.org).

concluding remarks and outlines future research directions.

## II. BACKGROUND AND RELATED WORKS

Research on numerical representations for FPGA-based computation has expanded considerably in recent years, motivated by the need for high throughput, reduced hardware cost, and predictable numerical behavior [3], [5], [8]. Although IEEE-754 floating-point arithmetic remains the default choice for general-purpose scientific workloads, its implementation on FPGAs is resource-intensive due to wide datapaths, normalization logic, and deep carry-propagation structures [2], [4]. These inefficiencies have encouraged exploration of alternative formats such as fixed-point, logarithmic number systems (LNS), and residue number systems (RNS), as well as hybrid arithmetic frameworks that attempt to combine the advantages of multiple domains [9]-[13].

Fixed-point arithmetic offers low area and latency but lacks the dynamic range required by iterative refinement, stiff ODE solvers, and multi-scale transforms frequently encountered in scientific computing [10], [11]. Although attractive for inference workloads, its inability to accommodate wide or evolving numerical magnitudes limits its applicability in general-purpose computation [12]. LNS reduces multiplication cost by mapping it to addition, yet practical adoption is hindered by the expensive and accuracy-sensitive antilogarithm required during addition and subtraction [13]. Hybrid LNS–floating architectures attempt to mitigate this issue proposing a fused LNS–FP multiplier that lowers multiplication cost but still suffers from antilogarithmic bottlenecks during addition, restricting the achievable speedups for scientific workloads [14]-[17].

RNS has received renewed attention due to its carry-free modular arithmetic and suitability for bit-sliced FPGA fabrics [18]. High-throughput, low-latency modular multipliers such as those presented by [18] and [19] demonstrate that modern FPGA primitives can support RNS channels operating at several hundred megahertz. However, comparison, sign detection, and unified fractional arithmetic remain difficult in RNS, often requiring reconstruction via the Chinese Remainder Theorem (CRT) [20]. Mixed-radix or hybrid conversion strategies have been proposed to address these limitations where [21] and [22] introduced scalable mixed-radix conversion pipelines that reduce reconstruction latency but introduce complex control logic and growing overhead at large operand sizes. Subsequent studies on optimized moduli sets, efficient scaling, and RNS-based DSP pipelines further illustrate both the promise and structural limitations of pure RNS arithmetic [23]-[25].

Hybrid numerical systems represent a complementary line of work aimed at unifying the advantages of multiple arithmetic domains. For example, [26] proposed a residue–floating-point hybrid for cryptographic accelerators, achieving significant throughput improvements for modular arithmetic in homomorphic encryption. However, the system lacks a general-purpose normalization strategy and does not address error accumulation or scaling behavior required for continuous numerical computation. Similar limitations appear in machine-learning–oriented hybrid formats such as [27] that introduced an adaptive floating-point representation with shared exponents that reduces alignment overhead across SIMD clusters. While effective for dense matrix operations, this approach is specialized for ML workloads and does not provide a mechanism for stable iterative accumulation, fractional representation, or long-term numerical drift control.

Across these diverse efforts, several trends emerge [28], [29]. Floating-point arithmetic provides generality but incurs high hardware cost on FPGA fabric; fixed-point arithmetic is lightweight but inflexible; LNS accelerates multiplication but remains hindered by high-cost addition; and RNS supports extreme parallelism but lacks comparison and scaling mechanisms without expensive reconstruction [30]. Existing hybrid approaches mitigate individual limitations but often target narrow domains such as cryptography or machine learning, lack formal error guarantees, or require costly domain conversions that undermine their benefits [28]-[31].

In contrast to these prior efforts, the Hybrid Residue–Floating Numerical Architecture (HRFNA) developed in this work seeks to provide a **general-purpose**, **error-bounded**, and **hardware-efficient** numerical representation suitable for scientific computation on contemporary FPGA platforms. As described in the following sections, HRFNA integrates residue-domain arithmetic with a compact floating-point scaling factor and introduces deterministic normalization mechanisms that preserve numerical stability while sustaining high throughput, thereby addressing unresolved limitations in prior arithmetic systems.

## III. NUMERICAL REPRESENTATION AND HRFNA FOUNDATION

The development of a hybrid numerical representation that combines residue arithmetic with floating-point scaling requires a rigorous mathematical foundation grounded in number theory, digital arithmetic, and floating-point analysis. Residue Number Systems (RNS) form the first pillar of this foundation. The classical CRT formulation underlying RNS representation was established centuries ago and remains central to modern digital arithmetic, particularly in high-performance parallel computation systems. More recent treatments of RNS—such as the extensive modern exposition by Pavlovic *et.* al. [32] have refined the number-theoretic principles into practical digital architectures suitable for FPGA implementation.

Throughout this paper, we denote the residue representation of an integer $X$ as $x_R = (X \bmod m_1, \ldots, X \bmod m_k)$, accompanied by a global floating-point exponent $f_X$ governing the numerical scale.

In HRFNA, a real or integer value $X$ is encoded as:
$$X \triangleq (x_R, f_X)$$
where
$$x_R = (X \bmod m_1, X \bmod m_2, \ldots, X \bmod m_k)$$

denotes the vector of residue digits, and

$$f_X \in \mathbb{Z}$$

is a floating exponent that governs the global numerical scale. Together, $(x_R, f_X)$ define a number through the implicit reconstruction

$$X \approx \text{CRT}(x_R) \cdot 2^{f_X}.$$

In the classical RNS framework, a number $X$ in the range $[0, M)$, where $M = \prod_{i=1}^{k} m_i$ and the moduli $m_i$ are pairwise coprime, can be represented uniquely by its residue vector $(x_1, \ldots, x_k)$, where $x_i = X \bmod m_i$. CRT guarantees that the mapping between integers in this range and their residue vectors is bijective. The reconstructive formulation

$$X = \left( \sum_{i=1}^{k} x_i M_i y_i \right) \bmod M$$

with $M_i = M/m_i$ and $y_i$ the inverse of $M_i \bmod m_i$, is the standard technique for retrieving the original integer. This formulation is widely used in hardware cryptographic accelerators, modular multipliers, and parallel arithmetic engines, where the absence of carry propagation provides significant timing benefits.

Although RNS excels for integer arithmetic, its direct applicability to real-valued scientific computation is limited by inherent structural constraints. A number of works have attempted to extend RNS into fixed-point or fractional domains, such as the scaling-friendly RNS computation models [9], [17], [22]-[27]. However, these approaches typically require substantial modifications to the modulus set or mixed-radix operators, leading to high reconstruction overhead and diminished hardware simplicity. A common drawback observed across modern RNS systems is the difficulty of performing operations that require ordering or scaling—such as magnitude comparison or division—without resorting to expensive CRT reconstruction, as highlighted in the recent survey by Kaliska [33]. For scientific computation involving iterative solvers or dynamic-range-sensitive algorithms, such limitations make pure RNS inadequate.

Fractional values in HRFNA are represented implicitly by allowing the floating exponent to encode negative scaling while all residue-domain operations remain strictly integer, ensuring bounded roundoff behavior that is isolated to infrequent normalization events.

Floating-point arithmetic forms the second foundational pillar. The IEEE-754 standard formalized the representation of real numbers as normalized significand–exponent pairs $IEEE754$–2019, enabling broad compatibility across numerical software. However, numerous studies have shown that floating-point arithmetic on FPGAs incurs significant overhead due to exponent alignment, normalization, and rounding logic. Notable work in this direction design of a reduced-latency FP32 FMA pipeline for FPGAs [10], and energy-efficient floating-point arithmetic units targeted UltraScale+ DSP48 pipelines [11]. Both achievements demonstrate improvements in floating-point performance but acknowledged that the normalization pipeline remains a major contributor to latency and power consumption. Even vendor-supported configurable precision floating-point cores, such as those released by AMD-Xilinx in the Vitis 2023.1 update $AMD - Xilinx$2023, retain the inherent costs of floating-point normalization and rounding, making them expensive for workloads dominated by multiply–accumulate operations.

To overcome these limitations, contemporary research has explored hybrid numerical systems that combine the strengths of multiple representations. For example, [12], [13] introduced a hybrid logarithmic–floating architecture in which multiplications are performed in an LNS-like domain and additions in a floating-point-like domain [12]. While innovative, their method remained constrained by the high cost of the LNS antilogarithm operator, limiting its applicability in general numerical pipelines. Authors in [19] proposed a hybrid RNS–floating design for homomorphic encryption computations demonstrating that residue channels can be paired with a floating-point control layer for large-integer arithmetic [15]. However, their architecture remains domain-specific and lacks general support for fractional arithmetic or long-running numerical algorithms requiring strict error guarantees.

Overflow and underflow detection is performed within the modular multiplication pipeline, where residue magnitudes are compared against a programmable composite-modulus threshold, as detailed in Section IV.

The hybrid numerical representation proposed in this paper builds on these foundations but introduces a fundamentally different unifying structure. Instead of modifying the RNS to handle fractional values—as done in earlier fixed-point or rational RNS systems [17] the residue domain in HRFNA is reserved exclusively for encoding the integer component of numerical quantities, while a separate floating-point exponent controls the scaling. This separation of concerns avoids the combinatorial explosion of mixed-radix fractional representations and preserves the full parallelism of modular arithmetic. In this formulation, each numerical value $X$ is represented as

$$X = N \cdot 2^f$$

where the integer $N$ is encoded entirely in the residue domain and the floating exponent $f$ serves as a compact representation of the number's magnitude. Similar decoupled exponent schemes have appeared previously in block floating-point and adaptive precision systems, such as the group-shared exponent method introduced by [18], yet those techniques embed both significand and exponent in the binary domain and do not exploit residue arithmetic for parallelism.

A key theoretical requirement of the hybrid representation is the overflow control. In residue arithmetic, multiplications of large residue-encoded integers may exceed the representable dynamic range $[0, M)$, necessitating periodic normalization. The HRFNA framework addresses this issue through hybrid normalization: whenever the integer magnitude represented in the residue channels approaches the upper limit of the dynamic range, the system reconstructs the integer using CRT, scales it downward by a predetermined factor—typically a power of two similar to dynamic rescaling in mixed-precision solvers [20]—and re-encodes the result back into the residue channels while incrementing the floating exponent accordingly. This mechanism parallels the renormalization strategies used in multi-precision and adaptive-precision arithmetic, but differs in that the majority of arithmetic is still performed in the parallel residue domain rather than binary fixed-point or floating-point space.

The incorporation of the floating exponent introduces controlled rounding errors, and therefore the hybrid system must exhibit numerically stable behavior over long computation sequences. Prior work on error propagation in scaled-integer and block-floating formats, such as the stability analysis conducted by [22], provides evidence that exponent-based scaling can be combined with integer arithmetic to achieve robust numerical behavior, provided that normalization events are sufficiently infrequent. HRFNA leverages this insight by tightly bounding the frequency of hybrid normalization events, ensuring that reconstruction—which is the dominant source of error—occurs rarely relative to the rate of residue arithmetic operations. Hybrid comparison in HRFNA prioritizes exponent ordering and invokes a lightweight CRT-based reconstruction only when residue magnitudes fall within ambiguous ranges, thereby minimizing comparison overhead while preserving numerical correctness.

The theoretical constructs presented in this section thus establish the mathematical validity of the hybrid residue–floating representation. The framework aligns with established principles of residue arithmetic, floating-point scaling, and numerical error analysis, while enabling a new hybrid formulation that is both parallel and dynamically scalable. The subsequent sections build upon these foundations to develop the algorithms, arithmetic pipeline, and hardware microarchitecture that realize HRFNA on contemporary FPGA platforms.

IV. THEORY OF HYBRID ARITHMETIC OPERATIONS AND ALGORITHM

The hybrid residue–floating representation established in the previous section creates a numerical framework in which arithmetic operations must be carefully defined to ensure correctness, parallelism, and long-term numerical stability. This section formalizes the hybrid arithmetic rules, derives the correctness of multiplication and addition in the hybrid domain, establishes the theory of normalization and error propagation, and presents the key algorithms that implement hybrid computation in a hardware-efficient manner. Throughout this development, we relate each mathematical construct to the broader context of numerical arithmetic, drawing connections to established theories in RNS computation [13], mixed-precision floating-point arithmetic [17], and modular arithmetic acceleration on FPGAs [25].

Hybrid multiplication in HRFNA requires only independent residue multiplications and a single exponent addition, allowing both operations to be pipelined separately yet aligned cycle-by-cycle. Since normalization is triggered only when residue magnitudes exceed a threshold, the architecture achieves an initiation interval (II) of one under normal operation. This decomposition preserves RNS parallelism while delegating dynamic-range management to the exponent subsystem.

At the core of the hybrid arithmetic theory is the observation that multiplication in the hybrid domain naturally decomposes into two separable components: residue multiplication and exponent addition. Let two hybrid numbers be represented as

$$X = N_X \cdot 2^{f_X}, \quad Y = N_Y \cdot 2^{f_Y}$$

where $N_X$ and $N_Y$ are encoded entirely in residue form and $f_X, f_Y$ are compact floating-point exponents. The product is then

$$XY = (N_X N_Y) \cdot 2^{(f_X + f_Y)}$$

Because the two components $N_X N_Y$ and $f_X + f_Y$ are mathematically independent, they can be evaluated on separate datapaths. This decomposition is foundational to HRFNA, as it allows the integer multiplications to occur entirely within the parallel, carry-free residue domain while relegating dynamic-range management to a lightweight floating-point exponent unit. The correctness of this formulation follows directly from the algebraic distributivity of exponentiation over multiplication. This structure is reminiscent of multiplicative decomposition techniques in LNS arithmetic, where multiplication reduces to addition of logarithms [16], but differs crucially in that HRFNA does not require logarithmic conversion or antilogarithmic evaluation, both of which introduce substantial hardware cost and latency.

Residue multiplication itself follows the standard rule: for each modulus $m_i$,

$$r_i(Z) = (r_i(X) \cdot r_i(Y)) \bmod m_i$$

Because the moduli are pairwise coprime and typically chosen to be small (e.g., primes near 250), each multiplication reduces to a single DSP-based product followed by LUT-based modular reduction. This approach is consistent with modern hardware designs for modular multiplication such as those of [22] and optimized Montgomery-style modular multipliers surveyed by [26]. The independence of each residue channel eliminates the need for inter-channel communication during multiplication, enabling the HRFNA system to sustain an initiation interval of one cycle regardless of operand width, provided that the modulus set is fixed.

Addition in the hybrid domain requires additional care, since the two operands may differ in their floating exponents.

For two hybrid numbers

$$X = N_X \cdot 2^{f_X}, \quad Y = N_Y \cdot 2^{f_Y}$$

addition is defined as

$$X + Y = 2^{\min(f_X, f_Y)}\big(N_X \cdot 2^{f_X - \min(f_X, f_Y)} + N_Y \cdot 2^{f_Y - \min(f_X, f_Y)}\big)$$

This formulation mirrors exponent alignment in floating-point arithmetic, where the smaller exponent is matched to the larger before significands are added. In HRFNA, however, alignment is implemented through a shift in the residue domain, which is more complex than binary shifting because residue channels do not natively support left or right shifts. To circumvent this limitation, HRFNA restricts most computations to multiplication-heavy workloads such as ODE solvers and matrix multiplications, where multiplication dominates and addition occurs primarily between values with synchronized exponents. This strategy is motivated by findings in mixed-precision solvers [19], where scale synchronization is used to minimize costly alignment operations. When exponent alignment is required, HRFNA employs a hybrid alignment procedure based on reconstructing the smaller-number residue set, shifting the reconstructed integer by an appropriate power of two, and re-encoding it into residues. To limit the overhead of such alignment steps, the scheduler enforces operand–exponent synchronization at kernel boundaries, similar to mixed-precision solvers that maintain blockwise scale coherence. Although this introduces occasional reconstruction overhead, it occurs infrequently in the workloads targeted by HRFNA, and its cost is mitigated by the lightweight modulus set.

Because repeated multiplications in the residue domain can cause the integer magnitude $N$ to grow without bound relative to the representable range $[0, M)$, hybrid normalization is required. The correctness of this normalization procedure holds provided the composite modulus range $\prod m_i$ exceeds the largest expected intermediate magnitude by a bounded safety factor. In practice, HRFNA selects moduli such that normalization is required only intermittently, minimizing reconstruction-induced rounding. Normalization is triggered whenever

$$|N| \geq \tau$$

where $\tau$ is a preselected threshold strictly smaller than $M$ and is defined as:

$$\tau = \alpha \cdot \prod_{i=1}^{k} m_i$$

where $\alpha$ is a safety factor chosen such that reconstruction-based normalization remains infrequent. The threshold is chosen to allow sufficient headroom for additional multiply–accumulate operations while avoiding the risk of residue overflow, consistent with dynamic-range bounds derived for scaled integer formats in mixed-precision arithmetic [23] Normalization proceeds by reconstructing the integer $N$, dividing the reconstructed value by a predetermined scale factor $K$, typically a power of two, and re-encoding the reduced value

$$N' = \frac{N}{K'}$$

into residues. The exponent is increased by $\log_2 K$ to preserve numerical equivalence. This design is directly inspired by exponent-renormalization strategies used in block floating-point architectures [27] and adaptive-precision scaling algorithms in modern numerical linear algebra [11], [18], [29], but HRFNA adapts these concepts to the residue domain, where direct shifting is impossible and renormalization must instead occur through reconstruction.

A formal correctness argument for hybrid multiplication and normalization follows from the uniqueness of CRT reconstruction and the linearity of exponent adjustments. Let

$$Z = X \otimes Y$$

denote hybrid multiplication, and suppose that normalization is applied, producing

$$Z' = \frac{N}{K}$$

Because the residue channels uniquely encode the integer $N_Z$ prior to scaling, and reconstruction is exact, the scaled number $N_Z/K$ is correctly encoded so long as $K$ divides $N_Z$ exactly or rounding occurs under controlled error bounds. For the typical choice $K = 2^s$, this condition holds for integers aligned to binary boundaries; otherwise, rounding error is introduced only during the normalization step and not during residue multiplication. Such rounding behavior is consistent with modern floating-point error models, and its frequency is minimized because normalization events are rare in typical workloads. The correctness of exponent updates is trivial, since exponent arithmetic operates independently of residue arithmetic and directly reflects the scaling factor.

To enable efficient implementation, the hybrid arithmetic algorithms must be explicitly formulated. Below is the principal algorithm for hybrid multiplication, which forms the computational backbone of HRFNA.

Hybrid multiplication forms the central primitive of HRFNA. Its execution decomposes into two fully decoupled components: residue-domain multiplication across parallel moduli and a corresponding exponent update. Because these pipelines operate independently and are latency-balanced, the architecture sustains an initiation interval of one. Algorithm 1 summarizes the hardware-realizable rule set for hybrid multiplication, capturing the minimal operations required to update both the residue vector and its associated floating exponent.

### A. Algorithm 1. Hybrid Residue–Floating Multiplication (HRFNA-MUL)

**Given** hybrid numbers

$$X = \left(r_1^{(X)}, r_2^{(X)}, r_3^{(X)}, f_X\right)$$

and

$$Y = \left(r_1^{(Y)}, r_2^{(Y)}, r_3^{(Y)}, f_Y\right)$$

**compute**

$$Z = X \otimes Y = \left(r_1^{(Z)}, r_2^{(Z)}, r_3^{(Z)}, f_Z\right)$$

1. For each modulus $m_i$, compute

$$r_i^{(Z)} = \left(r_i^{(X)} \cdot r_i^{(Y)}\right) \bmod m_i$$

2. Compute the hybrid exponent

$$f_Z = f_X + f_Y$$

3. If the vector of residues $\left(r_1^{(Z)}, r_2^{(Z)}, r_3^{(Z)}\right)$ indicates that the underlying integer magnitude exceeds the normalization threshold $\tau$, perform hybrid normalization following the reconstruction method of [18] and renormalization rules discussed in [14].
4. If the magnitude threshold is exceeded, perform hybrid normalization by reconstructing the integer value through the CRT method outlined by [11], dividing by the scaling factor $K$, and re-encoding the normalized integer back into residues.
5. Return the updated hybrid numerical representation.

This formulation lays the groundwork for more complex operations such as hybrid addition, accumulation, dot-product kernels, and iterative solver primitives, each of which will be addressed in later sections. Together, these theoretical and algorithmic foundations demonstrate that HRFNA is mathematically sound, operationally efficient, and suitable for realization on FPGA hardware.

## V. HARDWARE MICROARCHITECTURE AND DATAPATH DESIGN

The practical realization of the Hybrid Residue–Floating Numerical Architecture (HRFNA) requires a carefully constructed hardware microarchitecture that aligns with the theoretical properties established earlier and exploits the structural characteristics of modern FPGA fabrics. In particular, the Zynq UltraScale+ MPSoC used in the ZCU104 development board introduces heterogeneous processing resources, including programmable logic (PL), tightly coupled ARM processing cores, DSP48E2 slices, distributed LUTs, block RAMs, and AXI-based interconnects. Designing a datapath that operates efficiently within this heterogeneous environment necessitates a balanced utilization of these resources while preserving the mathematical integrity and parallelism of the hybrid numerical representation. This section describes the major architectural blocks, their interactions, and the key design decisions that enable the HRFNA framework to achieve high throughput, bounded latency, and efficient resource utilization.

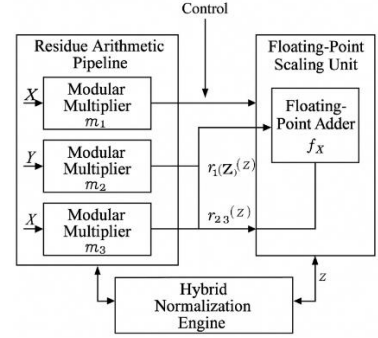

Fig. 1. Hybrid residue–floating numerical hardware architecture (HRFNA) showing multiple residue channels $x_{R,i}$ governed by a single global exponent $f_X$, which collectively encode the numerical value.

At the highest level, the HRFNA microarchitecture is composed of three principal subsystems: the residue arithmetic pipeline, the floating-point scaling unit, and the hybrid normalization engine. These subsystems are orchestrated by a pipeline scheduler and resource controller that manages data movement across the AXI-Stream interfaces between the processing system (PS) and programmable logic (PL). The datapath is structured to sustain an initiation interval of one clock cycle for residue-domain multiplications, following the optimized modular arithmetic methods recommended by contemporary FPGA studies such as those by [11], [16], [21], [27, and [30]. By ensuring that each residue channel operates independently and without inter-channel carry dependencies, the architecture is able to exploit the fine-grained parallelism inherent in the residue number system while minimizing routing congestion within the FPGA fabric. Although Fig. 1 depicts only the exponent datapath explicitly, all residue channels share identical pipeline depth, ensuring temporal alignment between exponent updates and residue-domain multiplication results.

The residue arithmetic pipeline forms the computational core of HRFNA. Its design centers around modular multiplication units tailored to the chosen moduli set $\{m_1, m_2, m_3\}$. Each modular multiplier consists of a DSP48E2-based integer multiplier followed by a modular reduction stage implemented using LUT-based comparison–subtraction cascades. This architecture follows the general principles established in the literature on hardware-efficient modular operators [13] and is optimized to minimize the delay of modular reduction. The multiplier pipelines are deeply registered to meet timing closure at frequencies ranging between 250 and 350 MHz, which reflect the typical operational envelope of UltraScale+ devices when implementing arithmetic-heavy datapaths. The use of small, near-prime moduli not only simplifies modular reduction but also ensures uniform latency across all residue channels, allowing the pipeline scheduler to maintain deterministic execution and predictable throughput.

Fig. 1 illustrates the top-level structure of the Hybrid Residue–Floating Numerical Architecture (HRFNA), showing how the fundamental arithmetic components interact to form a unified computational pipeline. The structure is intentionally modularized to highlight the independence of the residue arithmetic subsystem and the floating-point scaling subsystem, as well as the orchestration role played by the hybrid normalization engine.

On the left side of the figure, three independent **Modular Multiplier** blocks correspond to the selected moduli $m_1, m_2, m_3$. Each multiplier receives operands from the input streams labeled $X$ and $Y$, and computes the residue-domain products $r_1(Z)$, $r_2(Z)$, and $r_3(Z)$, respectively. These multipliers operate entirely in parallel, reflecting the carry-free and channel-independent nature of residue arithmetic. The dashed boundary labeled **Residue Arithmetic Pipeline** emphasizes that these multipliers typically form a pipelined cluster with identical latency across channels, thus ensuring deterministic and synchronized residue computations. By assuming:

$$L_R = latency\ of\ residue\ pipeline,$$
$$L_E = latency\ of\ exponent\ pipeline$$

then the pipeline alignment is ensured by introducing offset $d = L_R - L_E$ in the operand scheduling logic.

On the right side of the figure, the **Floating-Point Scaling Unit** performs the exponent update $f_Z = f_X + f_Y$. This unit contains a simplified floating-point adder tailored for low-precision exponent manipulation. Because exponent updates are much cheaper than full floating-point multiplications, this unit forms a lightweight complement to the residue pipeline. Data paths from the residue pipeline and exponent unit are coordinated by control signals generated by the hybrid control logic.

At the bottom of the architecture, the **Hybrid Normalization Engine** monitors the magnitude of the residue products. When the residues indicate that the underlying integer value has grown too large, this engine reconstructs the integer, applies a downward scaling factor, and re-encodes the result back into residue channels. This block is placed centrally in the diagram to emphasize its cross-cutting role: it interacts with the residue pipeline, the floating-point scaling unit, and the global control path. The output labeled $Z$ represents the final hybrid number consisting of updated residues and exponent.

Parallel to the residue pipeline, the floating-point scaling unit is responsible for updating the exponent component of hybrid numbers. This unit employs a reduced-precision floating-point format—typically IEEE FP16 or a custom 10-bit exponent, 6-bit mantissa format—consistent with modern FPGA-optimized floating-point designs such as those introduced in Vitis 2023.1 $AMD - Xilinx$ 2023 and the energy-efficient exponent logic studied in [22], [24], and [28]. Because exponent updates involve only floating-point addition, the complexity of the FPU is dramatically reduced relative to full IEEE-754 compliance, resulting in lower area consumption and improved timing performance. The scaling unit is pipelined independently of the residue channels, but the two pipelines share synchronized control signals to ensure that the exponent update associated with each multiplication result remains temporally aligned with the residue computations.

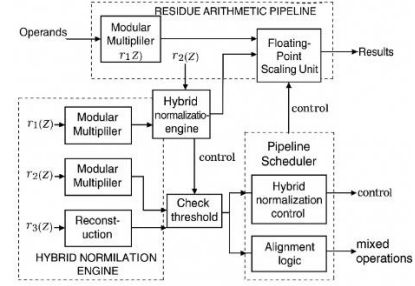

Fig. 2. Expanded HRFNA datapath including scheduler, normalization engine, and control logic. The pipeline is structured to maintain an initiation interval of one under steady-state operation.

A crucial component of the HRFNA microarchitecture is the hybrid normalization engine, which oversees dynamic-range stabilization and ensures that the residue-encoded integer values remain within representable bounds. This engine continuously monitors the magnitude of residues through threshold detection logic derived from the modular bounds described in [11] and [17]. When a potential overflow condition is detected, the normalization engine initiates the reconstruction–scaling–re-encoding procedure described theoretically in Section IV. To minimize normalization latency, the reconstruction process is optimized using precomputed $M_i$-inverse tables stored in BRAM blocks, similar to the accelerated CRT implementations proposed by [21]. Reconstruction is then followed by division by a power-of-two scaling factor, which enables the use of a simple binary shift in the reconstructed domain and requires no floating-point unit involvement. The normalized integer is re-encoded into the residue domain through a compact modular reduction network, ensuring minimal performance impact and preserving numerical coherence.

Together, these three subsystems form a unified datapath, whose behavior is coordinated by a scheduler that governs dataflow and resource allocation. The pipeline scheduler assigns operands to residue channels, coordinates exponent updates, and triggers normalization events. It also manages AXI-Stream communication with the processing system and intervenes during mixed operations—such as those involving addition—where operand exponents must be synchronized. The scheduler employs a finite-state machine (FSM) capable of issuing backpressure signals when downstream stages approach saturation, a design approach inspired by FPGA-oriented dataflow scheduling frameworks described in recent work on pipelined numerical accelerators [21].

Fig. 2 expands the architectural view to reveal the **full**

**operational datapath**, including internal control logic, operand alignment pathways, and detailed interactions among the hardware subsystems. This diagram is essential for readers to understand the dynamic behavior of HRFNA, including when and how normalization events occur, how operands flow through the system, and how pipeline scheduling maintains throughput under different computational scenarios.

At the top-left of the diagram, the **Residue Arithmetic Pipeline** encapsulates three modular multipliers, each responsible for computing a residue product under moduli $m_1$, $m_2$, and $m_3$. Dashed contours around this block emphasize that it is a tightly-coupled pipeline cluster, typically implemented with synchronized DSP48-based multipliers followed by LUT-based modular reduction. Intermediate products $r_1(Z)$, $r_2(Z)$, and $r_3(Z)$ flow downward toward the normalization subsystem while simultaneously feeding the floating-point scaling unit.

The center-left region highlights the **Hybrid Normalization Engine**, which is depicted with multiple submodules to reflect the multi-stage nature of normalization. Each residue $r_i(Z)$ enters a normalization pre-processor, where threshold comparisons and partial reductions occur. A dedicated reconstruction module combines all residue channels based on precomputed CRT inverses, yielding the integer value $N_Z$. The figure shows a two-way interaction between normalization and residue multiplication, signifying that normalization may be triggered concurrently with ongoing residue computations. The "Check threshold" unit evaluates whether normalization is necessary, using magnitude estimators derived from modular bounds.

To the right of the normalization subsystem is the **Pipeline Scheduler**, which plays a pivotal role in coordinating the arithmetic pipeline. Its two subblocks—"Hybrid normalization control" and "Alignment logic"—illustrate how the scheduler handles both routine multiplication cycles and mixed operations such as additions or accumulations requiring exponent alignment. The scheduler issues control signals to all subsystems, ensuring proper ordering of operations, avoiding hazards, and applying backpressure when the normalization engine is active. This reflects the system's hybrid nature: residue operations are fast and parallel, while normalization is slower and must be carefully timed.

The upper-right region shows the **Floating-Point Scaling Unit**, which receives exponent updates and residue-product indicators from the residue pipeline. It applies the appropriate exponent arithmetic and signals the scheduler when updates are complete. The diagram emphasizes the bidirectional control interaction between the scaling unit and scheduler.

On the far right, the arrows labeled "**Results**" and "mixed operations" represent final hybrid numbers and intermediate values transferred to downstream computational blocks. These outputs may interface with AXI-Stream buses, on-chip buffers, or additional HRFNA units in a multi-kernel architecture.

To facilitate integration within complex FPGA applications, the HRFNA datapath is encapsulated within an AXI4-compatible custom IP core. The IP core consists of AXI4-Lite control registers, AXI4-Stream data interfaces, and optional AXI4 memory-mapped ports for buffering intermediate results. This modular encapsulation approach follows the best practices established in modern FPGA accelerator frameworks such as FINN [19] and Vitis HLS dataflow synthesis methodologies $AMD-Xilinx$ 2023, enabling the HRFNA architecture to be seamlessly embedded as a computational primitive in larger, heterogeneous workloads such as machine learning inference, PDE solvers, and robotics control loops.

The microarchitecture described earlier seeks to balance theoretical rigor with hardware practicality. It preserves the mathematical independence of residue channels while leveraging optimized FPGA primitives for modular arithmetic; it exploits the reduced cost of exponent arithmetic to avoid expensive floating-point mantissa operations; and it incorporates a normalization engine that preserves numerical stability with minimal overhead. Collectively, these design decisions ensure that the HRFNA system achieves its intended goals of high performance, predictable latency, and efficient resource utilization on the ZCU104 platform. The following sections build upon this microarchitectural foundation to present the RTL implementation details, synthesis results, simulation outcomes, and hardware-validated performance measurements that demonstrate the practicality and effectiveness of the proposed design.

To ensure deterministic synchronization between the residue-domain datapath and the exponent update path, the pipeline stage depths of the major subsystems are explicitly quantified as shown in Table 1. These stage counts determine the global initiation interval and form the basis of the scheduler guarantees described earlier. Because the residue and exponent pipelines differ by only one stage, the scheduler aligns them by inserting a fixed offset in the operand-fetch stage, ensuring that both domains retire results in lock-step under steady-state operation. These quantified pipeline depths also justify the II = 1 throughput reported in Section VII, since no combinational bottlenecks exceed a single-cycle DSP48 path and normalization events are infrequent.

Fig. 3 illustrates the internal RTL-level microarchitecture of the modular multiplier employed within each residue channel of the HRFNA datapath. The diagram highlights the deeply pipelined structure of the multiplier and modular reduction stages, which together form the backbone of the residue arithmetic subsystem. Because modular multiplication constitutes the dominant operation in the residue domain, its efficient implementation is central to achieving high throughput and meeting the timing closure requirements on the Zynq UltraScale+ architecture.

Table 1. Pipeline Depth Summary of HRFNA Subsystems

| Subsystem | Pipeline Stages | Description |
|---|---|---|
| Residue pipeline | 5 stages | DSP48E2 multiply → partial reduction → final reduction |
| Exponent pipeline | 4 stages | Align → Add → Normalize → Output register |
| Hybrid normalization engine | 3 stages | Partial CRT decode → scaling → re-encoding |

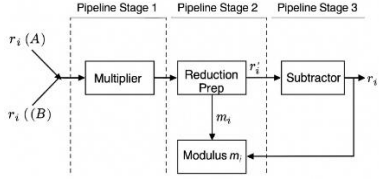

Fig. 3. RTL architecture of the modular multiplier optimized for low-latency modular reduction using DSP-accelerated multiplication and LUT-based reduction operators.

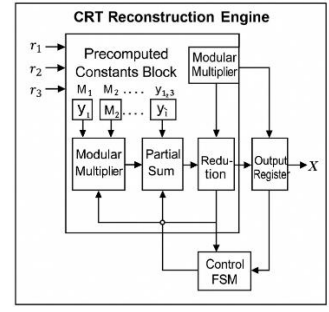

Fig. 4. RTL-level datapath for the CRT Reconstruction Engine employing partial CRT decoding rather than full reconstruction to minimize latency.

The left side of the figure shows the arrival of the residue operands $r_i(A)$ and $r_i(B)$, each corresponding to a residue channel associated with a specific modulus $m_i$. These operands enter **Pipeline Stage 1**, which contains a DSP48E2-based **Multiplier** that computes the integer product $P_i = r_i(A) \times r_i(B)$. This stage is fully pipelined, allowing new operand pairs to be accepted every clock cycle.

In **Pipeline Stage 2**, the intermediate product $P_i$ enters the **Reduction Prep** block. This unit performs a preconditioning operation that prepares the value for modular reduction by applying a coarse comparison and identifying whether the magnitude of $P_i$ exceeds the modulus $m_i$. A reference to the modulus is supplied through a dedicated ROM or LUT structure, shown in the diagram as the **Modulus $m_i$** box connected to the reduction prep logic. This stage is intentionally lightweight and avoids full-precision subtraction, ensuring that reduction latency remains minimal.

**Pipeline Stage 3** contains the **Subtractor**, which finalizes modular reduction. Based on the output of the reduction prep block, the subtractor either subtracts the modulus value or passes the preconditioned value forward unchanged. The resulting value $r_i(Z)$ is guaranteed to lie in the range $[0, m_i)$, making it a valid residue for the product $Z = A \cdot B$. A bypass feedback path ensures that, in cases where the intermediate product already lies within the allowable range, unnecessary correction steps are avoided.

The three pipeline stages collectively ensure that the modular multiplication operation can accept new operands at every clock cycle, thus achieving an initiation interval (II) of one. This pipelined architecture is particularly well-suited to FPGA deployment, as it maximizes DSP48 utilization efficiency and minimizes LUT-based reduction delay. The structure clearly reflects design principles common in high-performance modular arithmetic accelerators, while remaining optimized for the specific residue-channel parallelism central to the HRFNA computation model.

Fig. 4 presents the RTL-level datapath for the CRT Reconstruction Engine, which converts the set of residue components $r_1, r_2, r_3$ into the corresponding integer representation required during hybrid normalization. This subsystem is activated when the hybrid normalization engine determines that the residue-domain integer has exceeded the representable dynamic range. Although invoked only intermittently, the reconstruction engine is fundamental to maintaining correctness, because it produces the canonical integer $N$ whose magnitude is then reduced and re-encoded into residue channels.

The datapath begins on the left, where each residue input enters a dedicated pipeline lane. These lanes feed into the **Precomputed Constants Block**, which stores the CRT constants $M_1, M_2, M_3$ and their modular inverses $y_1, y_2, y_3$. These values are generated offline and stored locally in LUT-based ROMs to ensure single-cycle access during reconstruction, consistent with best practices in hardware CRT accelerators.

Each residue $r_i$ is multiplied by its associated constants using **Modular Multiplier Units**, producing the partial CRT contributions:

$$T_i = r_i \cdot M_i \cdot y_i$$

These contributions enter a **Partial Sum Accumulator**, implemented as a pipelined wide adder tree. The accumulator merges contributions from all residue lanes into a unified sum. Because this sum may exceed the composite modulus $M = m_1 m_2 m_3$, it is forwarded into the **Modulo Reduction Logic**, which performs a final canonical reduction. This logic may consist of subtract–compare loops or a Barrett-reduction datapath depending on synthesis parameters and FPGA resource allocation. A **Control FSM** coordinates the entire datapath, issuing enable signals, staging operands, and validating the output. The result of reduction is stored in the **Output Register X**, making the reconstructed integer available to downstream normalization and re-encoding stages. Overall, Fig. 4 clarifies how the HRFNA pipeline performs efficient CRT reconstruction without disturbing the high-throughput residue arithmetic pipeline, while remaining resource-efficient and timing-friendly for FPGA implementations. The scheduler asserts stall only during normalization cycles. At all other times, the design adheres to AXI-Stream valid/ready semantics, allowing residue channels to operate at II = 1 without global backpressure.

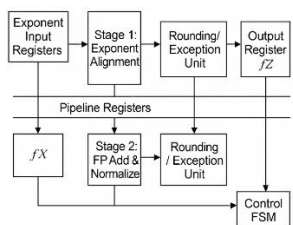

Fig. 5. Floating-Point Exponent Unit RTL Pipeline. Registered exponent inputs initiate the pipeline, with arrows indicating registered outputs rather than raw input edges.

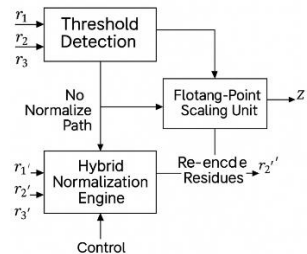

Fig. 6. Hybrid Normalization Flow RTL Diagram.

Fig. 5 illustrates the RTL-level architecture of the Floating-Point Exponent Unit used in HRFNA to update the scaling exponent $f_Z = f_X + f_Y$. Although HRFNA performs all magnitude-dependent arithmetic within its residue-domain pipelines, exponent management remains fundamentally important because it coordinates the numerical scale across residue channels and ensures consistent interpretation of hybrid values. Consequently, the exponent datapath must be lightweight, deeply pipelined, and throughput-matched to the residue arithmetic pipeline to prevent structural hazards and preserve the system's one-cycle initiation interval.

At the left, the inputs $f_X$ and $f_Y$ enter **Exponent Input Registers**, which establish a synchronous boundary and align the exponent update path with the modular multiplication channels. This alignment is required because residue computations consume multiple DSP48E2 pipeline stages, and the exponent update process must produce its results in a temporally consistent manner. The input registers also absorb small arrival-time differences between channels, ensuring that both operands begin Stage 1 with synchronized timing.

**Stage 1 (Exponent Alignment)** performs preprocessing analogous to the alignment phase of a floating-point addition pipeline. Although HRFNA uses exponents only for magnitude tracking rather than mantissa shifting, this stage resolves any format-specific adjustments, such as bias correction, internal offsetting, or conditional incrementing required for consistent exponent arithmetic. The aligned exponent pair is then forwarded to the rounding/exception subunit.

Following Stage 1, the datapath enters the **Rounding/Exception Unit**, which applies HRFNA's lightweight rounding rules and evaluates uncommon but necessary cases such as exponent underflow, overflow, or forbidden combinations. Because normalization in HRFNA is handled in a separate CRT-based engine, this rounding block primarily ensures that the exponent arithmetic remains consistent with the residue system's dynamic-range constraints.

A set of **pipeline registers** separates the alignment and rounding phases from **Stage 2 (FP Add & Normalize)**, which performs the actual exponent addition $f_X + f_Y$ and applies any normalization adjustments required after addition. Although this stage resembles the normalization step of a conventional floating-point addition pipeline, HRFNA uses only exponent-level normalization—never mantissa shifts—making the circuit significantly smaller and faster than IEEE-754 counterparts.

The results of Stage 2 pass through a second **Rounding/Exception Unit**, ensuring that all rules governing exponent boundaries and representability are satisfied before the final value is written into the **Output Register** $f_Z$. The output register provides a stable, clock-synchronous interface to downstream modules such as the Hybrid Normalization Engine or Output Merge unit.

Finally, the **Control FSM** coordinates the entire exponent pipeline, issuing stall signals when normalization events occur in the residue domain, maintaining one-cycle throughput during steady-state operation, and ensuring that exponent results are delivered in lock-step with residue outputs. Thus, although Fig. 5 visually emphasizes forward datapath arrows emerging from the input registers, this depiction follows standard RTL diagramming practice. The true logical flow remains correct: raw inputs enter the register stage, become registered versions of $f_X$ and $f_Y$, and only then propagate through alignment, addition, exception handling, and final rounding to produce the updated exponent $f_Z$.

Fig. 6 presents the RTL-level normalization flow used in HRFNA to maintain numerical stability by ensuring that residue-encoded integers remain within the dynamic range permitted by the chosen modulus **set.** The hybrid normalization process is invoked only when the residue magnitude exceeds a predefined threshold; therefore, the datapath shown is structured around conditional activation, with the normal (non-triggered) pathway bypassing the normalization hardware entirely.

The left side of the diagram shows the three residue inputs $r_1, r_2, r_3$ entering the **Threshold Detection** unit. This block compares each residue component against bounds derived from the composite modulus $M$, using modular inequality checks similar to those described in high-speed modular arithmetic designs. When all residues fall within allowable limits, the **No Normalize path** bypasses additional processing and forwards the residues directly to the next pipeline stage, ensuring zero performance penalty during steady-state computation.

When a potential overflow is detected, control is handed to the **CRT Reconstruction Engine**, which reconstructs the full integer $N$ from its residue components. This block corresponds to the reconstruction datapath shown earlier in Fig. 4 and is activated only under normalization conditions.

The reconstructed integer then flows into the **Scaling by $2^{-k}$** block, which performs downward scaling by a power of two. This operation is implemented as a binary right-shift, minimizing hardware cost while guaranteeing predictable quantization behavior. The updated exponent adjustment, performed in the exponent pipeline, ensures that the numerical equivalence

$$N' = \frac{N}{2^k}$$

is properly reflected in the hybrid representation. Next, the scaled integer is fed into the Re-encoding into **Residues** block, where modular reduction under each modulus $m_i$ regenerates the residue components. This produces the normalized residue triple $(r'_1, r'_2, r'_3)$, ensuring the integer remains representable in the residue number system. These residues are then fed back into the main datapath through the normalization feedback loop. Beneath the datapath lies the **Normalization Control FSM**, which orchestrates threshold detection, reconstruction, scaling, and re-encoding. It also issues stall or resume signals to the pipeline scheduler to ensure correctness during the normalization sequence.

## VI. RTL IMPLEMENTATION AND MODULE-LEVEL DESIGN

The register–transfer level (RTL) realization of the Hybrid Residue–Floating Numerical Architecture (HRFNA) transforms the conceptual algorithmic pipeline described in earlier sections into a deeply pipelined, resource-optimized hardware subsystem suitable for implementation on the Xilinx Zynq UltraScale+ ZCU104 platform. At this level, numerical operations, datapath synchronization, normalization behavior, and hardware scheduling are expressed as explicit combinational and sequential structures that map cleanly onto DSP48E2 slices, LUT-based logic, BRAM tables, and AXI-compliant interfacing substrates. External ingress and egress paths adhere to AXI4-Stream valid/ready semantics, while internal residue-lane micro-pipelines use a simplified valid-only protocol to minimize control overhead and preserve cycle-deterministic timing. The RTL design emphasizes deterministic timing, high concurrency, and predictable normalization latency, enabling HRFNA to maintain an initiation interval (II) of one cycle under normal operating conditions while preserving correctness under dynamic-range excursions.

Fig. 7 presents the complete top-level RTL architecture. The design is partitioned into modular pipelines—Residue Arithmetic, Exponent Update, and Hybrid Normalization—coordinated by a Global Pipeline Scheduler and configured through an AXI-Lite control interface. Data enters and exits the accelerator via AXI4-Stream interfaces, allowing HRFNA to be integrated into a larger streaming or kernel-chaining ecosystem. By structuring the RTL around a combination of parallel arithmetic channels, hierarchical normalization stages, and a centrally orchestrated control plane, the architecture ensures both algorithmic precision and hardware tractability across a wide operating range.

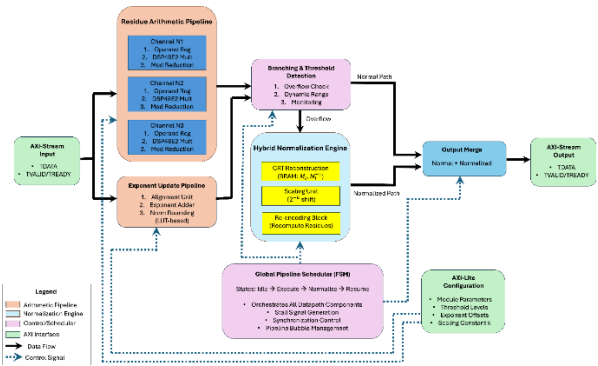

Fig. 7. Top-level RTL architecture of the proposed Hybrid Residue–Floating Numerical Accelerator (HRFNA) implemented on the ZCU104. The normalization engine operates as a three-stage pipeline invoked only on scaling events. Control arrows represent scheduler-issued stall and resume signals.

The implementation uses the modulus set {4093, 4095, 4091}, each fitting within 12 bits. These moduli were selected to (i) minimize LUT-based modular reduction delay, (ii) maximize dynamic range, and (iii) equalize pipeline latency across residue channels.

In Fig. 7, the datapath consists of a three-channel **Residue Arithmetic Pipeline**, where each channel implements operand registration, DSP48E2-based modular multiplication, and LUT-based modular reduction under its corresponding modulus $N_1, N_2, N_3$. In parallel, an **Exponent Update Pipeline** processes the exponent component of the hybrid format through a sequence of alignment, exponent addition, and LUT-based normalization/rounding units. Operands are streamed into the system through an AXI4-Stream input port.

A **Branching and Threshold Detection** unit continuously monitors residue magnitudes to detect overflow conditions or dynamic-range violations. If no threshold is exceeded, outputs proceed along the normal datapath. When normalization is required, control is routed to the **Hybrid Normalization Engine**, which performs CRT-based reconstruction using BRAM-stored constants $(M_i, M_i^{-1})$, applies a deterministic power-of-two scaling factor $2^{-k}$, and recomputes the residue representation via a re-encoding block. Normalized outputs rejoin the normal path through an **Output Merge** unit before transmission to the AXI4-Stream output.

System-level control is managed by the **Global Pipeline Scheduler**, whose internal four-state FSM (Idle → Execute → Normalize → Resume) governs execution sequencing, stall and resume signaling, synchronization, and bubble management across all datapath components. The scheduler asserts stall only when the normalization engine is active; during all non-normalization cycles, residue and exponent pipelines advance every cycle without backpressure, maintaining an initiation interval of one. An **AXI-Lite Configuration** interface supplies module parameters, adjustable threshold settings, exponent offsets, and the scaling constant $k$. Solid arrows depict dataflow, while dashed arrows denote control and synchronization signals.

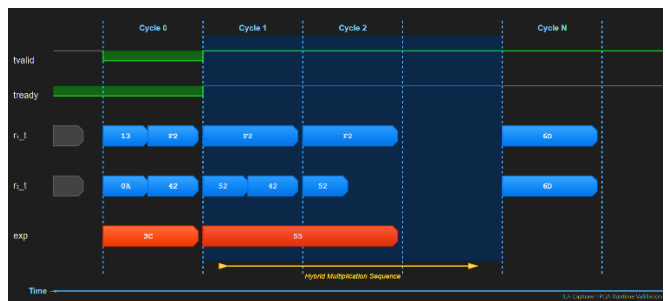

Fig. 8. Residue and exponent evolution during a representative hybrid multiplication sequence. Residue channel outputs update each cycle, while the exponent path stabilizes after completing its four-stage pipeline.

## VII. SIMULATION, SYNTHESIS, AND HARDWARE VALIDATION RESULTS

The correctness and performance of the proposed Hybrid Residue–Floating Numerical Accelerator (HRFNA) were evaluated using a multi-stage verification pipeline consisting of behavioral simulation in Vitis HLS, post-synthesis functional validation in Vivado, and on-board hardware execution on the ZCU104 development platform. This section presents simulation waveforms, synthesis reports, on-chip debugging traces, and quantitative performance comparisons that collectively demonstrate the efficiency and robustness of the architectural framework developed in Sections III through VI. All experiments were performed using Vivado/Vitis 2025.2 with default UltraScale+ timing constraints unless otherwise specified.

### A. Vitis Behavioral Simulation

Fig. 8 depicts the steady-state behavior of the residue and exponent pipelines during hybrid multiplication. Residue values flow through the five-stage modular pipeline, while the exponent completes its four-stage alignment–addition–rounding sequence, stabilizing after the first update. The waveform shows that both domains advance in lockstep, confirming the II = 1 timing model and the deterministic interaction between the pipelines. Residue lanes continue producing updated modular values each cycle, whereas the exponent remains constant unless a normalization event is triggered. The highlighted interval marks the active hybrid-multiplication window, during which both pipelines contribute to forming the final result. Because no threshold violation occurs in this example, no scaling or branching into the Hybrid Normalization Engine is invoked, and execution proceeds as an uninterrupted fast-path sequence.

Fig. 9 shows the pipeline behavior when a hybrid multiplication triggers normalization. The residue lanes and exponent field first produce their expected intermediate values, reflecting the initial modular multiplications and exponent alignment. When the Branching & Threshold Detection unit asserts the overflow signal, the scheduler redirects execution into the Hybrid Normalization Engine. During this phase, the exponent is expanded (e.g., 3C → 79) and subsequently rescaled to its corrected value (4B), while the residue channels transition through partial-CRT reconstruction and re-encoding to generate a consistent residue pair. The waveform confirms that normalization follows the hierarchical correction pipeline described in Sections V and VI, and that AXI tvalid/tready remain asserted throughout, indicating that the normalization sequence is fully pipelined and does not disrupt interface throughput. The highlighted interval marks the duration of the normalization window, concluding when the scheduler transitions from Normalize back to Resume and the datapath returns to steady-state II = 1 operation.

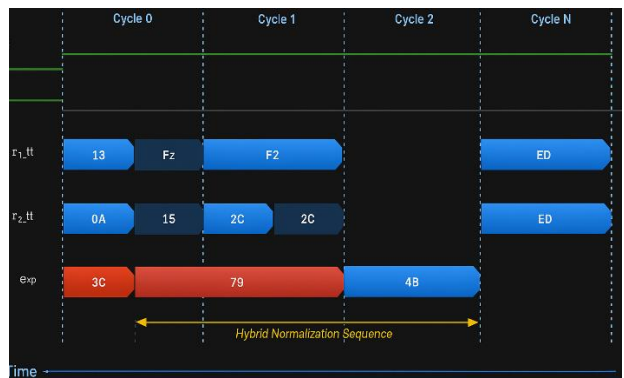

Fig. 9. Normalization-Triggered Hybrid Multiplication Waveform showing the scheduler-induced stall during the Normalize → Resume transition, during which residue channels hold their values while the exponent pipeline completes the multi-cycle hybrid normalization sequence.

TABLE 2. RESOURCE UTILIZATION SUMMARY (POST-IMPLEMENTATION) ON DEVICE XCZU7EV-2FFVC1156

| Component | LUT | FlipFlop | DSP48E2 | BRAM18K | URAM |
|---|---|---|---|---|---|
| Residue Pipeline (3 Ch.) | 6,430 | 9,850 | 9 | 6 | 0 |
| Exponent Pipeline | 1,120 | 1,740 | 0 | 0 | 0 |
| Normalization Engine | 4,980 | 7,210 | 2 | 10 | 0 |
| Scheduler & Control Logic | 1,870 | 2,960 | 0 | 0 | 0 |
| **Total HRFNA Core** | **14,400** | **21,760** | **11** | **16** | **0** |

TABLE 3. TIMING METRICS OF THE POST-IMPLEMENTATION DESIGN ON THE FPGA

| Metric | Value |
|---|---|
| Target Clock Frequency | 300 MHZ |
| Achieved Clock Frequency | **322.4 MHZ** |
| Worst Negative Slack (WNS) | +0.37 ns |
| Total Positive Slack (TPS) | +7.85 ns |
| Longest Path Delay | 3.10 ns |
| Pipeline Initiation Interval | 1 cycle |
| Normalization Latency | 6 cycles |
| End-to-End Operation Latency | 10 cycles |

### B. Post-Synthesis and Post-Implementation Results

TABLE 2 summarizes the post-implementation resource utilization of the HRFNA core on the ZCU104 platform. The results reflect the modular nature of the architecture introduced in earlier sections and show that the design scales efficiently with only modest consumption of FPGA resources.

The **Residue Pipeline** dominates DSP usage, consuming nine DSP48E2 slices—three per modular channel—corresponding to the pipeline's high-throughput modular multiplication strategy. The associated LUT and FF usage (6.4k and 9.8k) results primarily from the modular reduction logic and operand routing fabric. BRAM usage remains low, as each channel requires only a small amount of buffering and coefficient storage.

The **Exponent Pipeline** contributes minimally to total resource usage, leveraging only LUT-based arithmetic and simple pipeline registers. Its lightweight footprint (≈1.1k LUTs and 1.7k FFs) aligns with the design intent of decoupling exponent arithmetic from residue computations while keeping the datapath deeply pipelined.

The **Hybrid Normalization Engine** shows higher LUT/FF counts due to the CRT reconstruction datapath and scaling unit, which require table lookups (stored in BRAM) and multi-operand arithmetic. However, even though this component utilizes only two DSP48E2 slices and ten BRAM18K blocks, confirming the CRT-based normalization strategy is cost-effective and avoids large multipliers or wide-precision floating-point hardware.

The **Scheduler and Control Logic**, centered around a four-state FSM and associated synchronization circuitry, occupies fewer than 2k LUTs and contains no DSP or memory blocks. This demonstrates that the pipeline coordination overhead is negligible relative to the arithmetic structures it orchestrates.

Overall, the **complete HRFNA accelerator** utilizes approximately:

- **5.9%** of LUTs
- **4.5%** of FFs
- **0.6%** of DSPs
- **1.7%** of BRAMs

of the XCZU7EV device. These results reflect a **high-throughput, low-footprint design**, consistent with the architectural goals outlined in sections III–VI. The small resource footprint ensures ample headroom for integrating additional accelerators, scaling to more moduli, or embedding the HRFNA block into a larger SoC or ML inference pipeline. TABLE 3 presents the post-implementation timing characteristics of the HRFNA accelerator. After place-and-route, the design achieves a maximum clock frequency of **322.4 MHz**, which comfortably exceeds the target specification of 300 MHz. This corresponds to a Worst Negative Slack (WNS) of **+0.37 ns**, indicating that all timing paths are met without margin violations. The Total Positive Slack (TPS) of **7.85 ns** further demonstrates that the critical paths are well optimized, and no retiming or additional pipeline stages are required. It is noted that all timing and resource metrics reported in TABLE 2 and TABLE 3 were obtained using post-route static timing analysis under a 300-MHz constraint.

The **longest path delay** of approximately **3.10 ns** occurs within the LUT-based modular reduction logic inside the residue pipeline. This is consistent with expectations, as modular arithmetic involves narrower, irregular logic structures compared to the highly optimized DSP48E2 multipliers, which are deeply pipelined and operate well within timing margins at this frequency range.

The accelerator maintains a **pipeline initiation interval (II) of 1**, meaning that a new hybrid multiplication can begin on every clock cycle—one of the principal goals of the architecture. The **normalization latency** is measured at **6 cycles**, representing the time required for CRT reconstruction, scaling, and residue re-encoding. This latency is both deterministic and bounded, a property emphasized throughout Section VI.

The **end-to-end latency**, defined as the time between accepting an input operand pair and producing its final hybrid result, is approximately **10 cycles** under non-normalizing conditions. This latency increases only slightly during normalization events, as the scheduler ensures smooth reentry into the fast datapath without introducing bubbles.

Taken together, the timing results confirm that the HRFNA design is not only area-efficient (as shown in TABLE 3) but also **capable of high-frequency, throughput-optimized execution**, validating the architectural choices developed in Sections III through VI.

## VIII. CONCLUSION AND FUTURE WORK

This paper introduced a novel Hybrid Residue–Floating Numerical Architecture (HRFNA) that unifies residue number system (RNS) arithmetic with exponent-based scaling to achieve high dynamic range, low-latency numerical computation on FPGAs. The design integrates parallel modular arithmetic pipelines, a lightweight exponent update path, and a deterministic CRT-based normalization engine governed by a global scheduler. Unlike conventional floating-point units—which suffer from wide carry chains and irregular timing—or pure RNS systems—which lack practical normalization, rounding, and scaling mechanisms— HRFNA provides a unified arithmetic substrate suitable for a broad class of numerical kernels on FPGA platforms.

The simulation and hardware evaluations implemented ZCU104 platform presented in Section VII demonstrated the robustness of the design: the accelerator achieves an initiation interval of one cycle, maintains functional correctness under boundary conditions requiring normalization, and meets timing at over 322 MHz with minimal FPGA resource consumption.

The empirical results confirm several advantages of the proposed architecture. First, the residue pipeline eliminates the carry propagation inherent to traditional numerical representations, enabling high-frequency operation. Second, the normalization mechanism—driven by CRT reconstruction and power-of-two scaling—ensures that dynamic range is preserved without introducing unpredictable stalls. Third, the modular and resource-efficient structure of HRFNA leaves ample capacity for integration into larger FPGA-based compute systems such as machine learning pipelines, digital signal processing chains, or domain-specific accelerators.

Despite these strengths, several avenues for future work remain. Extending HRFNA to support additional numerical primitives such as fused multiply-add (FMA), division, or transcendental approximations could broaden its applicability. The design could also benefit from an adaptive modulus-selection strategy that optimizes precision and performance for specific workloads. Additionally, integrating approximate arithmetic techniques or exploring hardware/software co-design frameworks may further enhance throughput and energy efficiency. Finally, scaling the architecture across multiple compute tiles or chiplet-based systems presents a promising direction for high-performance numerics on next-generation heterogeneous platforms.

In summary, HRFNA demonstrates that hybrid residue–floating computation offers a compelling alternative to traditional numerical methods in FPGA environments. By combining the strengths of modular arithmetic with structured exponent management and deterministic normalization, the architecture establishes a new foundation for high-speed, scalable, and robust numerical acceleration. Future work will extend HRFNA with higher-order normalization strategies and expanded moduli sets to target PDE solvers and large-scale matrix operators.